\documentclass[nofootinbib, reprint, preprintnumbers, amsmath, amssymb, amsfonts, aps, pra, superscriptaddress, floatfix]{revtex4-2}
\usepackage[utf8]{inputenc}
\usepackage{mathtools}
\usepackage{graphicx}
\usepackage{dcolumn}
\usepackage{bm}
\usepackage{physics}
\usepackage{color}
\usepackage{braket}
\usepackage[normalem]{ulem}
\usepackage{enumerate}
\usepackage{enumitem}
\usepackage[colorlinks=true,linkcolor=blue,citecolor=blue,urlcolor =blue]{hyperref}
\usepackage{mathrsfs}
\usepackage{bbm}
\usepackage{siunitx}
\usepackage{xcolor}
\usepackage{listings}

\usepackage{comment}
\usepackage{lipsum}
\newcommand{\YM}[1]{\textcolor{black}{#1}}

\newcommand{\hk}[1]{\textcolor{black}{#1}}
\newcommand{\ty}[1]{\textcolor{black}{#1}}

\newcommand{\RS}[1]{\textcolor{black}{#1}}

\usepackage{CJKutf8}

\begin{document}
\title{Quantum Circuit Learning Using Non-Integrable System Dynamics}

\author{Ryutaro Sato}
\affiliation{Department of Electrical, Electronic, and Communication Engineering, Faculty of Science and Engineering, Chuo university, 1-13-27, Kasuga, Bunkyo-ku, Tokyo 112-8551, Japan}

\author{Yasuhiro Aota}
\affiliation{Department of Electrical, Electronic, and Communication Engineering, Faculty of Science and Engineering, Chuo university, 1-13-27, Kasuga, Bunkyo-ku, Tokyo 112-8551, Japan}

\author{Takaharu Yoishida}
\affiliation{Department of Electrical, Electronic, and Communication Engineering, Faculty of Science and Engineering, Chuo university, 1-13-27, Kasuga, Bunkyo-ku, Tokyo 112-8551, Japan}
\affiliation{Department of Physics, Tokyo University of Science,1-3 Kagurazaka, Shinjuku, Tokyo, 162-8601, Japan}

\author{Hideaki Kawaguchi}
\affiliation{Medical Research Center for Pre-Disease State (Mebyo) AI, Graduate School of Medicine, The University of Tokyo, 7-3-1, Hongo, Bunkyo-ku, Tokyo 113-0033, Japan}
\affiliation{Graduate School of Science and Technology, Keio University, Yokohama, Kanagawa 223-8522 Japan}

\author{Yuichiro Mori}
\affiliation{Global Research and Development Center for Business by Quantum-AI Technology (G-QuAT),
National Institute of Advanced Industrial Science and Technology (AIST),
1-1-1, Umezono, Tsukuba, Ibaraki 305-8568, Japan}

\author{Hiroki Kuji}
\affiliation{Department of Electrical, Electronic, and Communication Engineering, Faculty of Science and Engineering, Chuo university, 1-13-27, Kasuga, Bunkyo-ku, Tokyo 112-8551, Japan}
\affiliation{Department of Physics, Tokyo University of Science,1-3 Kagurazaka, Shinjuku, Tokyo, 162-8601, Japan}
 
\author{Yuichiro Matsuzaki}
\affiliation{Department of Electrical, Electronic, and Communication Engineering, Faculty of Science and Engineering, Chuo university, 1-13-27, Kasuga, Bunkyo-ku, Tokyo 112-8551, Japan}

\date{\today}


\begin{abstract}
Quantum machine learning is an approach that aims to 
\YM{improve the performance of}
machine learning methods by leveraging the properties of quantum computers. In quantum circuit learning (QCL), a supervised learning method that can be implemented using variational quantum algorithms (VQAs), the process of encoding input data into quantum states has been widely discussed for its 
\YM{important role} 
on the expressive power of learning models. In particular, the properties of the eigenvalues of the Hamiltonian used for encoding significantly influence model performance. Recent encoding methods have demonstrated that the expressive power of learning models can be enhanced by applying exponentially large magnetic fields proportional to the number of qubits. However, this approach poses a challenge as it requires exponentially increasing magnetic fields, which are impractical for implementation in large-scale systems.
\YM{Here, we propose a QCL method that leverages a non-integrable Hamiltonian for encoding, aiming to achieve both enhanced expressive power and practical feasibility.}
\YM{We find that the thermalization properties of non-integrable systems over long timescales, implying that the energy difference has a low probability to be degenerate, lead to an enhanced expressive power for QCL.}
Since the required magnetic field strength remains within a practical range, our approach to using the non-integrable system is suitable for large-scale quantum computers. 
Our results bridge the dynamics of non-integrable systems and the field of quantum machine learning, suggesting the potential for significant interdisciplinary contributions.

%
\end{abstract}

\maketitle
\section{Introduction}
\label{sec:intro}
In recent years, machine learning has undergone rapid development, with research being conducted from various perspectives \cite{carleo2017solving, rupp2012fast, broecker2017machine, ramakrishnan2015big, august2017using}. Supervised machine learning is a framework that trains models using datasets to predict appropriate outputs for unknown inputs \cite{testbishop2006pattern}. First, \textcolor{black}{a set of} \textcolor{black}{the} input data and \textcolor{black}{the} corresponding 
\textcolor{black}{output} data 
\textcolor{black}{is} prepared \textcolor{black}{as a teacher data}. 
The model processes the input data and outputs a predicted value. 
\textcolor{black}{The differences} between this prediction and the 
\textcolor{black}{teacher output} data 
\textcolor{black}{compose} a cost function, and the model parameters are adjusted to minimize this 
\textcolor{black}{cost function}.
Various optimization algorithms are used for this adjustment. By repeating this process, the model learns to produce predictions  that approach the correct data, eventually enabling it to make accurate predictions for unknown data.

Quantum machine learning \cite{schuld2015introduction,schuld2019quantum,carleo2019machine,perez2020data,biamonte2017quantum} is an idea that aims to enhance machine learning performance by leveraging the properties of quantum computers. 
Quantum computers have been studied for their potential \YM{for speedup.}
However, quantum error correction is essential for quantum computers to outperform classical ones in large-scale problems \cite{gottesman2002introduction}. Implementing quantum error correction is technically challenging and expected to take a considerable amount of time to achieve. On the other hand, noisy intermediate-scale quantum (NISQ) computers, which 
\YM{does not need} error correction and have a relatively small number of qubits, are anticipated to become available in the near future \cite{bharti2022noisy, preskill2018quantum}. Variational Quantum Algorithms (VQAs) \cite{cerezo2021variational}, which combine classical and quantum computing, are particularly promising for NISQ devices due to their reliance on shallow quantum circuits. VQAs use parameterized quantum gates to construct quantum circuits, optimizing these parameters with classical computers.

Quantum Circuit Learning (QCL) is one method that uses quantum computers to build learning models and is categorized as a VQA algorithm \cite{mitarai2018quantum,schuld2019quantum}. Specifically, it can be applied to supervised machine learning. The computation is assigned to a quantum circuit, while parameter updates are performed on a classical computer. Classical data can be encoded into quantum states using qubit rotation angles and input into quantum circuits. The quantum circuit parameters are iteratively adjusted by introducing optimization methods commonly used in neural networks, and the optimized circuit generates outputs that closely approximate the target values.


\YM{There are studies about}
encoding input data from classical to quantum data, \YM{where} its impact on the expressive power of quantum learning models \YM{is analyzed} \cite{schuld2021effect, mori2024expressive}. 
The input data \RS{$x_i$, which is one of the elements in the training dataset $(x_i, y_i)_{i=1}^L$,} is encoded into the input state $e^{-ix_i\hat H}\ket{0\dots0}$ using a quantum gate $e^{-ix_i\hat H}$ and an initial state $\ket{0\dots0}$, where $\hat{H}$ \hk{represents an}
arbitrary Hamiltonian with eigenvalues $E_n (n=1, 2, \dots,2^N)$
, \hk{and} $N$ denotes the number of qubits. A variational quantum circuit $\hat{U}(\boldsymbol\theta)$ with updated parameters \hk{$\bm{\theta}$} acts on the input state\hk{.} 
\hk{This} results in the output state $\ket{\psi}=\hat{U}(\boldsymbol\theta)e^{-ix_i\hat H}\ket{0\dots0}$. 
The learning model $f_{\boldsymbol\theta}(x)$ is defined as the expectation value of some observable \hk{$\hat{M}$} with respect to the output state. 
\hk{Specifically, }the learning model can be expressed as:
\begin{equation}
     f_{\boldsymbol \theta}(x)=\bra{0\dots0}e^{ix\hat H}\hat U^{\dagger}(\boldsymbol \theta)\hat{M}\hat U(\boldsymbol \theta)e^{-ix\hat H}\ket{0\dots0}\hk{.} \label{eq:1.1}
\end{equation}
The learning model can also be represented as a Fourier series:
\begin{equation}
    f_{\boldsymbol\theta}(x)=\sum_{\omega\in\Omega}c_\omega(\boldsymbol \theta)e^{-i\omega x}, \label{eq:1.2}
\end{equation}
where $\Omega$ 
\textcolor{black}{represents} the frequency spectrum (the set of energy differences $E_n-E_m$), $\omega$ denotes a frequency component, and $c_\omega(\boldsymbol\theta)$ is the Fourier coefficient. The frequency spectrum \hk{$\Omega$} 
determines the number of Fourier components in the function $f_{\boldsymbol\theta}(x)$.
\YM{More specifically,} $\Omega$ represents the types of functions \textcolor{black}{that the quantum model} 
\textcolor{black}{$\hat{U}(\boldsymbol{\theta})e^{-ix_{i}\hat{H}}$} 
can 
\textcolor{black}{provide}, 
while $c_\omega(\boldsymbol\theta)$ 
\hk{are}
the controllable parameters of the model, which are adjusted during learning. Consequently, this encoding process influences the expressive power of the learning model. 

Various encoding methods in quantum machine learning have been analyzed through Fourier component analysis. A conventional method involves rotating \hk{to} all qubits uniformly using a Hamiltonian $\hat {H}_1=\sum_{i=1}^N\hat{Y}_i$, where \hk{$\hat X,\hat Y,\hat Z$ are the Pauli operators}
and $\hat Y_i$ represents the $\hat Y$ operator acting on the $i$-th qubit. 
\YM{This corresponds to uniform rotation on all qubits}. 
\hk{In this case, the number of distinct eigenvalues is $G=N+1$, and the number of distinct energy differences (frequency components) is $K \equiv|\Omega|=2N+1$, which corresponds to the number of unique pairs of energy differences $E_m-E_n$~\cite{schuld2021effect}.}
Another method 
\hk{utilizes}
an exponentially weighted Hamiltonian $\hat H_2 =\sum_{i=1}^N 3^i \hat Y^{(i)}$ \cite{PhysRevA.107.012422}. In this case, \hk{the number of distinct eigenvalues grows exponentially with the number of qubits,} $G=2^N$.
Compared 
\hk{to} the uniform rotation method, the energy difference degeneracy is significantly reduced, resulting in $K=3^N-1$. This indicates 
improved expressive power for the model. However, 
\hk{implementing this method requires applying} exponentially large magnetic fields 
as the number of qubits increases,
\YM{making this approach impractical.}
\YM{Moreover,} the theoretical limit for 
\hk{the number of distinct frequencies (assuming no degeneracy)} is $K=4^N-2^N+1$, which 
\hk{has} not yet achieved \hk{by existing method}.

\hk{In quantum mechanics,} integrable and non-integrable systems 
are crucial concepts for understanding the properties of quantum many-body systems. Integrable systems have sufficient \hk{number of} conserved quantities, and the level spacing distribution follows a Poisson distribution \cite{bender1998real}. On the other hand, non-integrable systems exhibit complex behavior due to a lack of conserved quantities, and the level spacing distribution generally follows a Wigner-Dyson distribution \cite{atas2013distribution, berry1977level, bohigas1984characterization}. This distribution is characterized by the phenomenon of "level repulsion
"\hk{,} where adjacent energy levels tend to repel each other, reducing energy degeneracy. Consequently, non-integrable systems typically 
\hk{thermalize} over 
\hk{long timescale} \cite{d2016quantum}.

The influence 
in \textcolor{black}{the} Hilbert space on learning performance has \textcolor{black}{already} been investigated \cite{sakurai2024simple}. 
\hk{Furthermore, } 
the dynamics of quantum many-body systems and their correlation with learning performance have been reported \cite{PhysRevA.108.042609}. However, the direct relationship between the dynamics of quantum many-body systems and quantum machine learning is non-trivial.

Here, we propose a method to enhance the expressive power of quantum learning models by using non-integrable Hamiltonians for data encoding. 
\YM{We find that the thermalization properties of non-integrable systems over long timescales, implying that the energy difference has a low probability to be degenerate, lead to an enhanced expressive power for QCL.}
Due to the level repulsion effect, energy degeneracy is minimized. 
These lead to an increased number of accessible Fourier components in the quantum learning model. Furthermore, this encoding method does not 
\hk{require} exponentially large coefficients like in the Hamiltonian, making it comparatively more straightforward to implement in systems with a large number of qubits.
\YM{By using numerical simulation, we show that our approach 
\hk{exhibit} better performance than the previous approaches for QCL.}
\YM{Our results bridge the dynamics of non-integrable systems and the field of quantum machine learning, 
which makes interdisciplinary contributions.}

\section{Quantum Circuit Learning as a NISQ Algorithm}\label{sec2}
\begin{figure}
    \centering
    \includegraphics[width = 9cm]{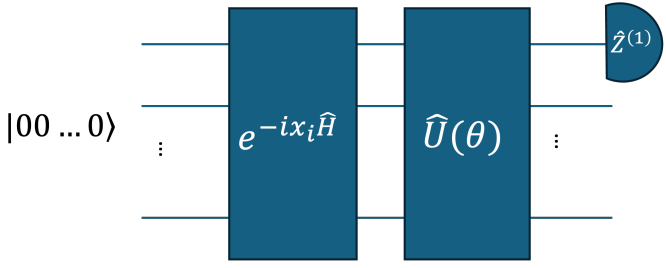}
    \caption{Four Kerr \hk{nonlinear} resonators are capacitively coupled \hk{to} a Josephson junction mode. The resonant frequencies of the Kerr non-linear resonators are far detuned from that of the Josephson junction mode. The \hk{nonlinearity} of the Josephson junction induces an effective four-body interaction \hk{among} the Kerr non-linear resonators.
    }
    \label{circuit}
\end{figure}

We review quantum circuit learning (QCL), which is one of the NISQ algorithms\cite{mitarai2018quantum}. This is a type of supervised machine learning
\hk{where} a training dataset $(x_i, y_i)_{i=1}^L$ is provided, where $L$ is the number of data in the dataset. It is assumed that there exists a \hk{target} function $\tilde{f}$, and there is a hidden relationship in the dataset between $x$ and $y$ such that $y=\tilde{f}(x)$.
The 
\hk{goal} of supervised learning is to discover $\tilde f$ from the training dataset. 
\hk{To this end}, we define a \YM{parametrized}
model function $f_{\theta}$ \hk{,}\YM{where $\theta$ denotes parameters},  and optimize \YM{the function} using the training data to 
\hk{approximate} $\tilde{f}$ \YM{by adjusting $\theta$}.
\YM{In QCL, we adjust a network composed of \hk{quantum} circuits, which is similar to neural networks.}
\hk{Specifically}, quantum gates 
are used to encode \hk{both} the input data $\boldsymbol x=(x_1,...,x_L)$ and the learnable parameters $\boldsymbol \theta=(\theta_1,...,\theta_M)$. By measuring the circuit multiple times, we estimate the expectation value of a specific observable, which we denote as the quantum model function $f_{\boldsymbol \theta}(x)$ 
, and this serves as our learning model based on quantum computing.

\hk{The cost function $L_{\mathrm c}(\boldsymbol \theta)$ evaluate the deviation between the model prediction and the teacher data. Here, we adopt the mean squared error as the cost function and aim to minimize it by optimizing the parameters $\boldsymbol\theta$.}
We first prepare the training dataset {$(x_i, y_i)$}$_{i=1}^N$ and encode the input value $x_i$ into a quantum state \YM{which is an initial state $\ket{00...0}$.} 
Using a Hamiltonian $\hat H$, we obtain the input state $e^{-ix_i\hat{H}}\ket{00...0}$.
\YM{To construct the variational quantum circuit, we apply a \hk{parametrized} unitary operator 
$\hat U(\boldsymbol \theta)$, 
\hk{resulting in the} output state $\hat U(\boldsymbol \theta)e^{-ix_i\hat{H}}\ket{00...0}$.}
The expectation value of the observable $\hat Z^{(1)}$ acting on the first qubit in the output state $\hat U(\boldsymbol \theta)e^{-ix_iH}\ket{00...0}$ is denoted as the output of $f_{\boldsymbol \theta}(x)$. Thus, we have
\begin{equation}
    f_{\boldsymbol \theta}(x)=\bra{00...0}e^{ix\hat{H}}\hat U^{\dagger}(\boldsymbol \theta)\hat{Z}\hat U(\boldsymbol \theta)e^{-ix\hat{H}}\ket{00...0}\label{eq.3}
\end{equation}
Letting the number of qubits be $N$ and inserting the identity operator, we obtain 
\begin{equation}
\begin{split}
    f_{\boldsymbol \theta}(x)= & \sum_{n=1}^{2^N}\sum_{m=1}^{2^N}e^{-i(E_n-E_m)x}\\
    & \langle 0\dots0\ket{E_n}\bra{E_n}\hat U^{\dagger}(\boldsymbol \theta) 
   \hat{Z}\hat U(\boldsymbol \theta)\ket{E_m}\langle E_m\ket{0\dots0}\label{eq.4}
\end{split}
\end{equation}
Here, $\ket{E_n}$ and $\ket{E_m}$ are the eigenstates of the Hamiltonian $\hat H$, and $E_n$ and $E_m$ are the corresponding eigenvalues. Using 
\hk{this} model \hk{function} $f_{\boldsymbol \theta}(x)$ 
\hk{and training data}, we define the cost function, which is the mean squared error:
\begin{equation}
    L_{\mathrm c}(\boldsymbol \theta) = \frac{1}{2}\sum_{i=1}^L(f_{\boldsymbol \theta}(x_i)-y_i)^2\label{eq.5}
\end{equation}
Methods such as the Nelder-Mead method are used for the minimization of the cost function.

We introduce the construction of $\hat U(\boldsymbol \theta)$. First, we perform time evolution based on the transverse magnetic field Ising model. 
\YM{The Hamiltonian is given as}
\begin{equation}
    \hat{H}= \sum_{i=1}^{N}a_i\hat{X}_i + \sum_{i=1}^{N} \sum_{j=1}^{i-1}J_{ij}\hat{Z}_i\hat{Z}_j\hk{,} \label{eq.6}
\end{equation}
\YM{where }
$a_i$ is the strength of the transverse field, and $J_{ij}$ denotes the strength of the interaction \hk{between the $i$th and $j$th qubits}.
We define a unitary operator for the data encoding as follows.
\begin{equation}
    \hat U_{\mathrm{rand}}=e^{-it\hat{H}} \label{eq.7}
\end{equation}
\YM{Also,} we utilize the $x$-axis rotation gate and the $z$ -axis rotation gate defined by
\begin{equation}
\hat{R}^X(\theta) = \begin{pmatrix}
\cos\frac{\theta}{2} & -i\sin\frac{\theta}{2} \\
-i\sin\frac{\theta}{2} & \cos\frac{\theta}{2}
\end{pmatrix}
\label{eq.8}
\end{equation}
and
\begin{equation}
\hat{R}^{Z}(\phi) = \begin{pmatrix}
e^{-i\frac{\phi}{2}} & 0 \\
0 & e^{i\frac{\phi}{2}}
\end{pmatrix}
\label{eq.9}
\end{equation}
We embed the parameters $\boldsymbol \theta$ as combinations of single-qubit gates defined by
\begin{equation}
  \hat U_{\mathrm{rot}}(\boldsymbol \theta_{i1}^{(d)}, \boldsymbol \theta_{i2}^{(d)}, \boldsymbol \theta_{i3}^{(d)})=\hat R_i^X(\boldsymbol \theta_{i1}^{(d)}) \hat R_i^Z(\boldsymbol \theta_{i2}^{(d)}) \hat R_i^X(\boldsymbol \theta_{i3}^{(d)})  \label{eq.10}
\end{equation}
Here, $d$ represents the layer (depth) of $\hat U_{\mathrm{rand}}$ and $\hat U_{\mathrm{rot}}$, and $i$ indicates the $i$ -th qubit being acted upon. Using the operators $\hat U_{\mathrm{rot}}(\boldsymbol \theta)$ from Eq.~\eqref{eq.7}\eqref{eq.10}, we construct $\hat U(\boldsymbol \theta)$:
\begin{equation}
    \hat U(\boldsymbol \theta) = \prod_{k=1}^d(\prod_{i=1}^N  \hat U_{\mathrm{rot}}(\boldsymbol \theta_i^{(d)})\hat U_{\mathrm{rand}}) \label{eq.11}
\end{equation} 
Here, the parameters $\boldsymbol \theta$ 
\YM{denotes} a vector of dimension $3dN$.


\section{Frequency Components and Representational Capacity of the Learning Model}

We review the 
\YM{expressive power}
\hk{of} the learning model when encoding input data $x_i$ into the input state $e^{ix_i\hat H}\ket{00...0}$. From Eq.~\eqref{eq.4}, we have $\omega=E_m-E_n$ and $c_\omega=\langle 00...0\ket{E_n}\bra{E_n}\hat U^{\dagger}(\boldsymbol \theta)\hat{Z}\hat U(\boldsymbol \theta)\ket{E_m}\langle E_m\ket{00...0}$. Defining the frequency spectrum $\Omega$ , we find that Eq.~\eqref{eq.4} can be expressed as $f_{\boldsymbol \theta}(x)=\sum_{\omega\in\Omega}c_\omega(\boldsymbol \theta)e^{i\omega x}$. Thus, if we consider $\omega$ as the frequency components and $c_\omega$ as the Fourier coefficients, we see that the quantum learning model $f_{\boldsymbol \theta}(x)$ in 
\hk{QCL} can be naturally represented as a Fourier series corresponding to $\Omega$. The set $\Omega$ determines the types of functions that the learning model can represent. In other words, if the frequency components
\YM{are}
degenerate, the variety of functions that the quantum learning model can represent is reduced. Therefore, it is desirable to choose a Hamiltonian $\hat H$ in such a way that the frequency components do not coincide as much as possible during the encoding process.  
In particular, consider a non-resonant condition in which $E_n-E_m=E_k-E_l$ holds only when $n=k$ and $m=l$. Satisfaction of this condition is ideal for enhancing the expressive power of the quantum learning circuit (QLC).

\YM{First,} we introduce a conventional encoding method using uniform rotation angles\cite{schuld2021effect}. Let $\hat Y^{(i)}$ denote the Pauli operator acting on the $i$-th qubit. When encoding the input $x_i$ into the input state $e^{-ix_iH}\ket{00...0}$ using the Hamiltonian \YM{described by}
\begin{equation}
    \hat{H}_1=\sum_{i=1}^{N}\hat{Y}_i \label{eq.12},
\end{equation}
we note that the eigenvalues of $\hat Y^{(i)}$ are $\pm1$.
\YM{This encoding procedure corresponds to a rotation of all qubits with a uniform angle.}
\YM{Let us denote the number of eigenvalues denoted as $G$, and we obtain}
\begin{equation}
    G=N+1 \label{eq.13}
\end{equation}
Additionally,
\YM{let us denote the number of non-degenerate energy differences of $\omega=E_m-E_n$ as $K$, and we obtain}
\begin{equation}
\begin{cases}
K = 2 & \text{(for } N = 1\text{)}, \\
K = 2N+1 & \text{(for } N > 1\text{)}. \label{eq.14}
\end{cases}
\end{equation}
Thus, under the encoding by $\hat H_1$, we can see that the variety of functions accessible to the learning model is approximately doubled compared to the number of qubits.

Next, we describe a recently proposed encoding method that utilizes exponential functions. We use the Hamiltonian 
\begin{equation}
    \hat{H}_2=\sum_{i=1}^{N}3^i\hat Y_i \label{eq.15}
\end{equation}
to encode the input $x_i$ into the input state $e^{-ixH}\ket{00...0}$. The number of eigenvalues $G$ is given by
\begin{equation}
    G=2^N \label{eq.16}
\end{equation}
and the number $K$ of frequency components is given by
\begin{equation}
    K=3^N-1 \label{eq.17}
\end{equation}
Comparing Eq.~\eqref{eq.14} and Eq.~\eqref{eq.17}, we can see that the number of frequency components in Eq.~\eqref{eq.17} increases with respect to the number of qubits. In other words, encoding to use an exponential function results in fewer degenerate frequency components compared to uniform rotation angle encoding, thereby increasing the variety of functions that the learning model can access. However, the encoding method that uses exponential functions necessitates the application of exponentially large magnetic fields, which makes practical implementation on actual machines difficult.

\section{Data Encoding Using the Dynamics of Non-Integrable Systems}
In 
\ty{this section}, we describe how the dynamics of non-integrable systems can be utilized for data encoding.
It is known that the Hamiltonians of non-integrable systems possess little to no symmetry, which significantly reduces the degeneracy of energy eigenvalues and leads to level repulsion between energy levels.
Furthermore, \ty{since} the non-resonance condition is closely related to thermalization phenomena, and 
\ty{almost all non-integrable systems are known to thermalize}, it is expected that the non-resonance condition is naturally satisfied in such 
\ty{, as we will discuss in detail in Appendix \ref{appendix}.}
Therefore, using the dynamics of non-integrable systems for data uploading is expected to enhance the expressive power of quantum circuits.



\YM{For the reasons described above,}
we propose to use the dynamics of the Hamiltonian of non-integrable systems for data encoding to achieve a quantum circuit learning model with high 
\ty{expressive power}. By leveraging the property that energy eigenvalues and their differences tend to avoid degeneracy, we expect that the Fourier components of the quantum model $f_{\boldsymbol \theta}(x)$ after Fourier transformation will increase exponentially.l
Moreover, unlike previous studies, our approach does not require the application of exponentially large magnetic fields, making it a more practical method. While there are prior examples that use the dynamics of non-integrable systems in the Ansatz circuits for learning, there are no examples yet that utilize the dynamics of non-integrable systems for data encoding in quantum circuit learning\cite{xiong2023fundamental, tangpanitanon2020expressibility}.

We adopt the following Hamiltonian that employs the dynamics of non-integrable systems:
\begin{equation}
\begin{split}
    \hat{H}_3 =& \sum_{i=1}^{N}B_i^X\hat X_i + \sum_{i=1}^{N}B_i^Y\hat Y_i + \sum_{i=1}^{N}B_i^Z\hat Z_i \\
    &+\sum_{i,j=1}^{N}J_{ij}(\hat X_i\hat X_j +\hat Y_i\hat Y_j + \Delta \hat Z_i\hat Z_j)\ty{,} \label{eq.18}
\end{split}
\end{equation}
\ty{where $\Delta$ is the anisotropy operator and set as $\Delta=0.73$ in this paper.}
Using this \ty{model}, we encode the input $x_i$ into the input state $e^{-ix_i\hat{H}}\ket{00...0}$. Here, the values of $B_i^{X}$, $B_i^{Y}$, $B_i^{Z}$ are randomly selected from a uniform distribution ranged as $[-1, 1]$, while the values of $J_{ij}$ are randomly selected from a uniform distribution ranged as $[-3, 3]$. In this case, the number of non-degenerate frequency components $K$ is expected to be given by 
\begin{equation}
    K=4^N-2^N+1 \label{eq.19}
\end{equation}
for $N$ qubits, since degeneracy is not anticipated. This represents the maximum number of frequency components that are theoretically allowed for $N$ qubits\cite{PhysRevA.107.012422}. \YM{In this case}, for $N=1, 2, 3, 4$, we have $K=3, 13, 57, 241$, respectively. From Section \ref{sec2}, it can be noted that the encoding using $\hat H_1$ and $\hat H_2$ yields for $N=1, 2, 3, 4$ the respective values of $K$ as follows: $\hat H_1$ gives $K=2, 5, 7, 9$ and $\hat H_2$ gives $K=3, 13, 57, 241$.
Encoding using the dynamics of non-integrable systems is expected to provide more frequency components relative to the number of qubits. Consequently, 
\YM{our proposal could} suggest an improvement in the representational capacity of the learning model, leading to the potential for higher performance in learning.

\section{Numerical Calculation Results}
\begin{figure}
    \centering
    \includegraphics[width = 9cm]{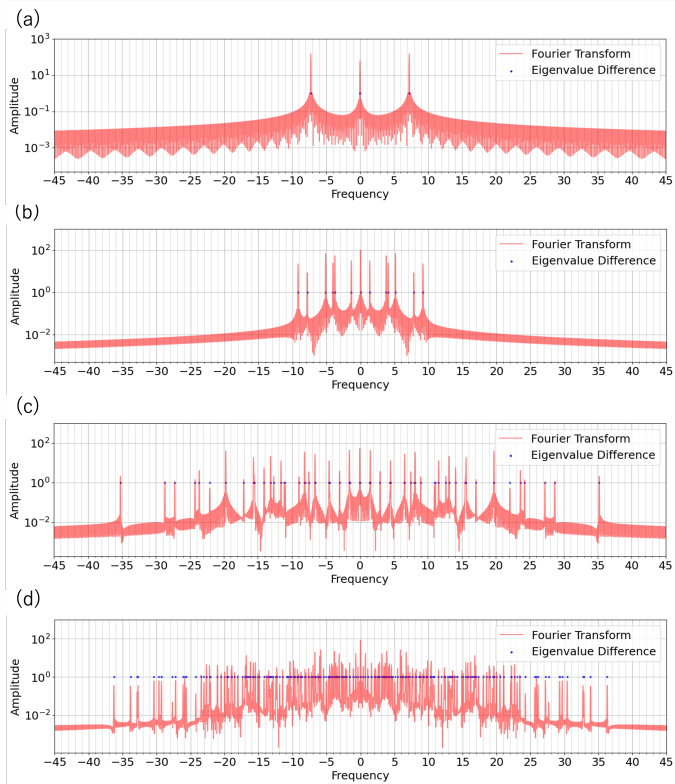}
    \caption{The results of the Fourier analysis of the learned model when data encoding is performed using the dynamics of a non-integrable system are shown. For comparison, the differences in the eigenvalues of $\hat{H}$ are also plotted. (a) Fourier components for the 1-qubit case. (b) Fourier components for the 2-qubit case. (c) Fourier components for the 3-qubit case. (d) Fourier components for the 4-qubit case.
    }
    \label{fig.zu1}
\end{figure}

\begin{figure*}
    \centering
    \includegraphics[width = 17cm]{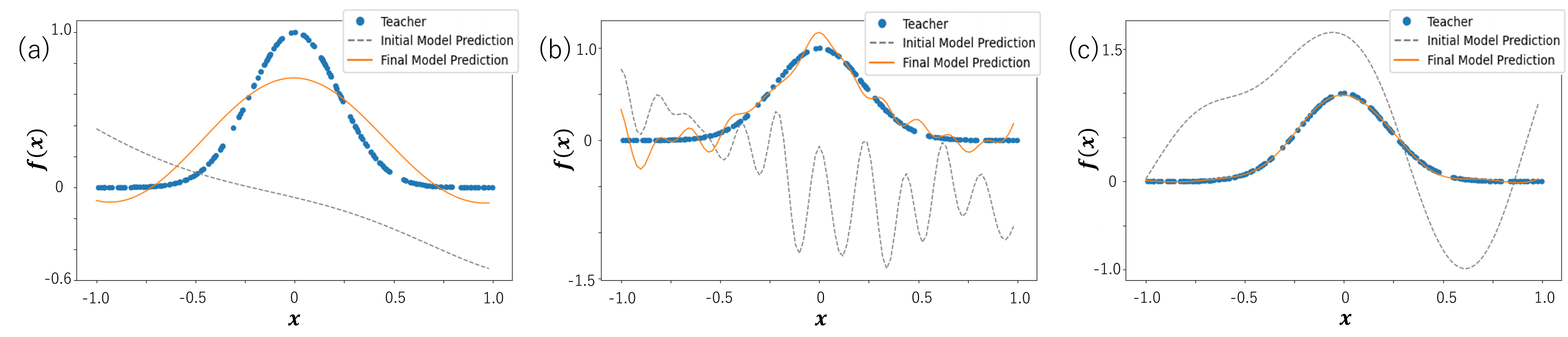}
    \caption{The learning results when the Gaussian function $\tilde{f}(x) = e^{-10x^2}$ is used as the target function to predict. The input includes both $\tilde{f}(x)$ and $f_{\boldsymbol{\theta}}(x)$, and the number of qubits is 4. (a) Learning results with encoding using uniform rotation angles. The cost function value at the initial parameters $\boldsymbol{\theta_0}$ is $3.11948 \times 10^{-1}$, and the cost function value at the optimized parameters $\boldsymbol{\theta_{\mathrm{opt}}}$ is $1.85351 \times 10^{-2}$. (b) Learning results with encoding using exponential functions. The cost function value at the initial parameters $\boldsymbol{\theta_0}$ is $1.00671$, and the cost function value at the optimized parameters $\boldsymbol{\theta_{\mathrm{opt}}}$ is $8.08237 \times 10^{-3}$. (c) Learning results with encoding using the dynamics of a non-integrable system. The cost function value at the initial parameters $\boldsymbol{\theta_0}$ is $4.76918 \times 10^{-1}$, and the cost function value at the optimized parameters $\boldsymbol{\theta_{\mathrm{opt}}}$ is $1.32745 \times 10^{-4}$.}
    \label{fig.zu2}
\end{figure*}

\begin{figure*}
    \centering
    \includegraphics[width = 17cm]{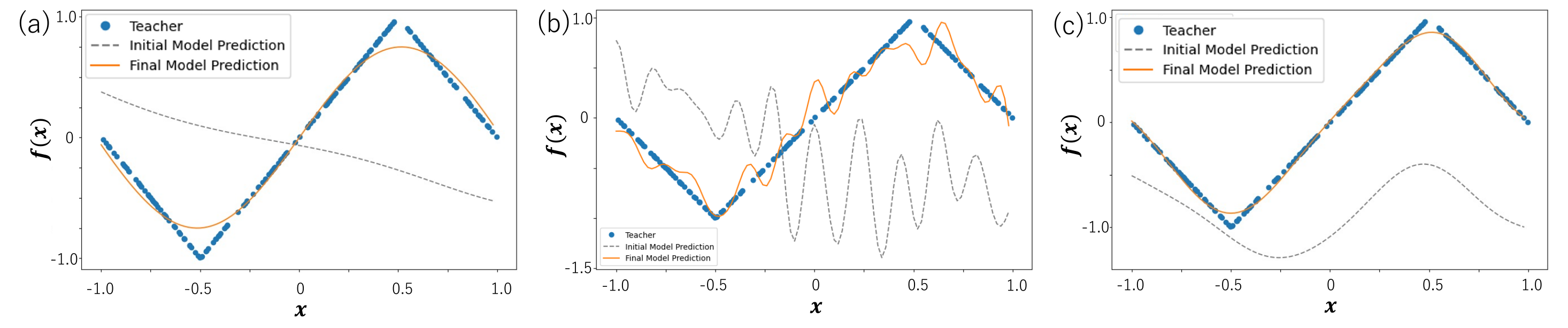}
    \caption{The learning results when the triangular wave with period 2 and amplitude 1 is used as the target function $\tilde{f}(x)$. The input includes both $\tilde{f}(x)$ and $f_{\boldsymbol{\theta}}(x)$, and the number of qubits is 4. (a) Learning results with encoding using uniform rotation angles. The cost function value at the initial parameters $\boldsymbol{\theta_0}$ is $5.71575 \times 10^{-1}$, and the cost function value at the optimized parameters $\boldsymbol{\theta_{\mathrm{opt}}}$ is $7.26426 \times 10^{-3}$. (b) Learning results with encoding using exponential functions. The cost function value at the initial parameters $\boldsymbol{\theta_0}$ is $1.00671$, and the cost function value at the optimized parameters $\boldsymbol{\theta_{\mathrm{opt}}}$ is $1.48816 \times 10^{-2}$. (c) Learning results with encoding using the dynamics of a non-integrable system. The cost function value at the initial parameters $\boldsymbol{\theta_0}$ is $1.48105$, and the cost function value at the optimized parameters $\boldsymbol{\theta_{\mathrm{opt}}}$ is $7.82698 \times 10^{-4}$.
    }
    \label{fig.zu3}
\end{figure*}

Using $\hat H_3$, we analyze the frequency components of the learning model 
\begin{eqnarray}
     f_\theta(x)=\bra{00...0}e^{ix\hat{H}_3}\hat U^{\dagger}(\theta)\hat{Z}^{(0)}\hat U(\theta)e^{-ix\hat{H}_3}\ket{00...0} \label{eq.18}
\end{eqnarray}
numerically through the discrete Fourier transform (DFT). The results are plotted in 
\ty{Fig.} \ref{fig.zu1}. For 1 to 4 qubits, it was confirmed that the number of peaks that appeared as a result of the DFT matches the theoretical maximum value given by Eq.~\eqref{eq.19}. Therefore, the number of frequency components is larger compared to conventional methods. 

\YM{Furthermore,} we compare the learning performance of the models \YM{where we encode the data by using}
the Hamiltonians $\hat{H}_1$, $\hat{H}_2$, and $\hat{H}_3$.
\YM{We use the Nelder-Mead method provided by SciPy library for optimization.}
The implementation of each algorithm is done using Python, and for the Nelder-Mead method, we utilize the SciPy library\cite{abcdefg, virtanen2020scipy}.
In Eq.~\eqref{eq.6} and Eq.~\eqref{eq.11}, $a_i$ and $J_{ij}$ are chosen from a uniform distribution between -1 and 1, with $d=3$, and the initial parameter values $\theta_0$ are randomly selected based on a uniform distribution from 0 to $2\pi$. 
\YM{First, we define the function that the learning model should predict 
a Gaussian function $\tilde{f}_\theta(x)=e^{-10x^2}$.}
The learning results are shown in 
\ty{Fig.} \ref{fig.zu2}. 
\YM{Second,} we present the learning results when the function that the learning model should predict is a triangular wave with a period of 2 and an amplitude of 1. \YM{We show the results} in 
\ty{Fig.} \ref{fig.zu3}. In both cases, compared to the encoding methods using uniform rotation angles and exponential functions, it is clear that our method \YM{accurately}
reproduces the function to be predicted with the highest accuracy. Moreover, a comparison of the cost function values after training also shows that our method 
\YM{provides} the lowest values, indicating that our approach is the most suitable for learning among these methods.

\section{conclusion}\label{relabel}
\YM{In conclusion,} we propose a 
data encoding method utilizing the dynamics of non-integrable systems \YM{for the data encoding} in quantum circuit learning.
By leveraging the characteristics of the Hamiltonian of non-integrable systems, particularly 
the suppression of degeneracy \ty{and resonance of energy eigenvalues}, it is expected that the Fourier components of the quantum model will increase exponentially, thereby enhancing the 
\textcolor{black}{expressive power} of the learning model.
The theoretical maximum of the Fourier components for a quantum model with $N$ qubits is known to be $4^N-2^N+1$, \YM{which the previous approaches cannot achieve.}
We confirmed through numerical calculations that our method achieves this theoretical maximum up to $4$ qubits. 
Additionally, since our proposed method does not require the application of exponentially strong magnetic fields that 
\ty{were necessary in the previous work}, it is anticipated to be a more practical approach. These results pave the way for new avenues in quantum circuit learning and are expected to provide an important \YM{bridge between quantum machine learning and quantum many body physics.}

\begin{acknowledgments}
R.S. acknowledges a helpful discussion with Miho Osanai and Miku Ishizaki.
This work is supported by JSPS KAKENHI (Grant Number 23H04390), JST Moonshot (Grant Number JPMJMS226C), CREST (JPMJCR23I5), and Presto JST (JPMJPR245B).

\appendix 
\section{Relationship between the thermalization and non-reonant conditions}\label{appendix}

\YM{In this section, we explain how the thermalization of quantum systems based on the eigenstate thermalization hypothesis (ETH) is related to the non-resonance condition of the energy eigenvalues of the system's Hamiltonian \cite{PhysRevA.43.2046, PhysRevE.50.888, riddell2024no}.}
\YM{In particular, we discuss the mechanism by which the expectation value of an observable relaxes to a steady state under the long-time dynamics governed by a unitary dynamics of the Hamiltonian.}
We consider a quantum system with the Hamiltonian given by $\hat{H} = \sum_n E_n \ket{E_n}\bra{E_n}$ and the initial state as $\ket{\psi(0)} = \sum_i c_i \ket{E_i}$.
\YM{Here, $c_i=\langle E_i|\psi (0)\rangle$, where $E_i$ are energy eigenvalues and $\ket{E_i}$ are the corresponding energy eigenstates.}
The time-evolved state is $\ket{\psi(t)} = e^{-i\hat{H}t} \ket{\psi(0)}$, and the expectation value of an observable $\hat{A}$ is given as follows
\begin{eqnarray}
    \langle \hat A(t) \rangle &=& \sum_{i,j} c_i^* c_j e^{-i(E_j - E_i)t} A_{ij}  
\end{eqnarray}
where $A_{ij} = \bra{E_i} \hat{A} \ket{E_j}$ denotes a transition matrix element.
We assume the following conditions:
\begin{itemize}
  \item Non-degeneracy:\\ $E_\alpha = E_\beta \Rightarrow \alpha = \beta$
  \item Non-resonance:\\ $E_\alpha - E_\beta = E_\gamma - E_\delta \Rightarrow (\alpha = \gamma,\ \beta = \delta)$
\end{itemize}
By using the non-degeneracy condition, the long-time average is given as
\begin{eqnarray}
\overline{\langle\hat A(t) \rangle} &=& \lim_{T \to \infty} \frac{1}{T} \int_0^T dt\, \sum_{i,j} c_i^* c_j e^{i(E_i - E_j)t}A_{ij}\\ &=& \sum_i |c_i|^2 A_{ii}. 
\end{eqnarray}
\textcolor{black}{Then, assuming the (diagonal) ETH, which is formulated as $\bra{E_i}A\ket{E_i}=\operatorname{Tr}[\rho_{\mathrm{MC}}A]$, we obtain
\begin{equation}
    \sum_i |c_i|^2 A_{ii}. \approx \operatorname{Tr}[\hat \rho_{\mathrm{MC}}(E)\hat A]
\end{equation}
where $\rho_{\mathrm{MC}}$ is the microcanonical ensemble, and $E$ represents a macroscopic energy and we assume that the majority of eigenstates concentrate in the energy window $[E-\Delta E,E]$ with $\Delta E$ is the energy width scales as $o(N)$
}

\YM{If the expectation value of an observable relaxes to the thermal expectation value under long-time dynamics governed by the Hamiltonian,}
\YM{then temporal fluctuations should vanish. Indeed, when the non-resonance condition is satisfied, temporal fluctuations in the long-time average of observables vanish in the thermodynamic limit, as shown below.}
\YM{The time fluctuation $\sigma_t^2$ is evaluated as follows:}
\begin{eqnarray}
    \sigma_t^2 &=& \overline{(\langle \hat A(t) \rangle - \overline{\langle \hat A \rangle})^2} \\
    &=& \lim_{T \to \infty} \frac{1}{T}\int_0^T dt\,
\sum_{i \neq j, k \neq l} c_i^* c_j c_k^* c_l \nonumber \\&&e^{i(E_i - E_j + E_k - E_l)t} A_{ij} A_{kl} \label{manyterms} \\
&=& \sum_{i \ne j} |c_i|^2 |c_j|^2 |A_{ij}|^2 \label{nonresonantuse} \\
&\leq& \left[ \max_{i \neq j} |A_{ij}| \right]^2
\end{eqnarray}
In deriving Eq.~\eqref{nonresonantuse} from Eq.~\eqref{manyterms}, the non-resonance condition is used.
According to the ETH, the off-diagonal elements $A_{ij}$ become extremely small in the thermodynamic limit. Thus, we obtain:
$\sigma_t^2 \leq \left[ \max_{i \neq j} |A_{ij}| \right]^2 \to 0$
This shows that the expectation value of the observable is close to its equilibrium value in the long-time limit.
\YM{In other words, this effectively shows the relaxation to a steady state.}
\YM{Although a rigorous proof is lacking, it is expected that nonintegrable quantum dynamics satisfy the non-resonance condition and exhibit thermalization as described above.}
\YM{Conversely, if the non-resonance condition is violated and there exist many pairs such that $E_\alpha - E_\beta = E_\gamma - E_\delta$ with $\alpha \neq \gamma$ or $\beta \neq \delta$,}
\YM{then the number of non-zero terms in Eq.~\eqref{manyterms} increases. In such cases, it becomes difficult for the system to relax to a steady state with vanishing temporal fluctuations.}

\end{acknowledgments}

\bibliography{ref}

\end{document}